\documentclass[11pt]{article}
\usepackage{xcolor}
\usepackage[english]{babel}
\usepackage[T1]{fontenc}
\usepackage[utf8]{inputenc}
\usepackage{authblk}
\usepackage{mathtools}
\usepackage{slashed}
\usepackage{amsmath,amssymb,mathrsfs}
\usepackage{amsfonts}
\usepackage{enumitem}
\usepackage{graphicx,color,xcolor}
\usepackage{cite}
\usepackage[multiple]{footmisc}
\usepackage{subcaption}
\usepackage{hyperref}
\hypersetup{
	colorlinks=true,
	linkcolor=blue,
	filecolor=magenta,
	urlcolor=cyan,
}
\usepackage{cleveref}

\usepackage{fontawesome5}
\usepackage{old-arrows}
\usepackage{adforn}
\usepackage{dsfont}

\numberwithin{equation}{section}

\definecolor{refcol}{rgb}{0.9,0.1,0.1}
\hypersetup{colorlinks=true,linkcolor=blue,citecolor=refcol,urlcolor=cyan,linktocpage}
\usepackage[left=2.5cm,right=2.5cm,top=2.5cm,bottom=2.5cm]{geometry}

\newcommand{\dow}{\partial}
\newcommand{\cL}{\mathcal{L}}
\newcommand{\cor}[2]{\big<#1|#2\big>}
\newcommand{\dR}{\mathrm{d}}
\newcommand{\Functions}{\mathscr{C}^\infty}
\newcommand{\cH}{\mathcal{H}}
\newcommand{\cQ}{\mathcal{Q}}
\newcommand{\g}{\mathfrak{g}}
\newcommand{\M}{\mathscr{M}}
\newcommand{\R}{\mathbb{R}}

\newcommand{\sst}{\scriptscriptstyle}
\newcommand{\smf}{\sf\scriptscriptstyle}

\begin{document}
	\begin{titlepage}
		\thispagestyle{empty}
		
		\title{
			{\huge\bf Reframing classical mechanics:
            An AKSZ sigma model perspective}\\ \hfill
		}
		
		\bigskip\bigskip\bigskip\bigskip\bigskip
		\vfill
		
		\author{
			{\bf Thomas Basile$^a$}\thanks{{\tt thomas.basile@umons.ac.be}},
			{\bf Nicolas Boulanger$^a$}\thanks{{\tt nicolas.boulanger@umons.ac.be}},
			{\bf Arghya Chattopadhyay$^{a,b}$}\thanks{{\tt arghya.chattopadhyay@upr.edu}}
			\smallskip\hfill\\      	
			\small{	
				${}^a${\it Service de Physique de l'Univers, Champs et Gravitation,Université de Mons}\\
				{\it 20 Place du Parc, 7000 Mons, Belgium}\\
				\smallskip\hfill\\
				${}^b${\it Physics Department, University of Puerto Rico Mayag\"uez}\\
				{\it Puerto Rico 00682, USA}\\
				\smallskip\vspace{1cm}
			}
		}

		\bigskip\bigskip\bigskip\bigskip

		\date{
			\begin{quote}
				\centerline{{\bf Abstract}}
				{\small
The path-integral re-formulation due to E. Gozzi, M. Regini, M. Reuter and W. D. Thacker of Koopman and von Neumann's original operator formulation of a classical Hamiltonian system on a symplectic manifold $M$ is identified as a gauge slice of a one-dimensional Alexandrov--Kontsevich--Schwarz--Zaboronsky sigma model with target $T^\ast(T[1]M\times \mathbb{R}[1])$.
				}
			\end{quote}
		}
		

	
\end{titlepage}

\thispagestyle{empty}\maketitle\vfill\eject

\baselineskip=18pt

\newpage

\tableofcontents

\section{Introduction}

In the early days of quantum mechanics, before the arrival
of the path-integral formalism, Koopman and von Neumann (KvN) 
\cite{Koopman:1931, von_Neumann:1932i, von_Neumann:1932ii}, 
motivated by improving their understanding of classical 
ergodicity, reformulated classical mechanics on a symplectic manifold $M$ as quantum mechanics on $T^\ast M$ 
with Hamiltonian linear in momenta and complex wave function
on $M$ given by the square root of the classical density 
function modulo the phase, whose fate remains an interesting 
open problem in KvN mechanics. 
Over half a century later, an equivalent path-integral 
formulation was given by E. Gozzi, M. Reuter,
and W. D. Thacker (GRT) \cite{Gozzi:1989bf, Gozzi:1991di}.

On the other hand, the Alexandrov--Kontsevich--Schwarz--Zaboronsky (AKSZ) formalism 
\cite{Alexandrov:1995kv}
provides a natural framework for the deformation quantization
of graded geometries appearing in the context of symplectic mechanics and gauge 
field theories. It was initially applied for the quantization of topological 
systems, but is also suitable to the description of dynamical systems with 
local degrees of freedom, provided these systems are described in terms 
of an exterior, Cartan-integrable system, 
see e.g. \cite{Grigoriev:2006tt,Boulanger:2012bj, Grigoriev:2019ojp} for discussions.

In this paper, we will show that the AKSZ formalism enables us to reframe 
in a very natural way
the path-integral formulation of classical mechanics starting from the 
operatorial formulation by 
Koopman and von Neumann (KvN) \cite{Koopman:1931, von_Neumann:1932i, von_Neumann:1932ii}, 
and further developed in \cite{Gozzi:1989bf, Gozzi:1991di}.
For that purpose, we will be studying the dynamics of a classical system whose 
phase space corresponds to the symplectic manifold $(M,\omega)$ and focus on its 
cotangent bundle $T^\ast M$, which is also symplectic. 
We will further subject $T^\ast M$ to a system of first class constraints that will provide us with a natural 
extension of the Koopman--von Neumann (KvN) reformulation of classical mechanics on $M$. More precisely, as we shall see, 
the link between KvN and AKSZ consists of the fact that the re-writing 
due to a series of work by Gozzi, Reuter and Thacker (GRT) \cite{Gozzi:1989bf, Gozzi:1991di} (and thereafter by Gozzi and Regini \cite{Gozzi:2000sf}) of the pull-back operation 
on $M$ along a symplectomorphism generated by a Hamiltonian vector field during a 
fixed time $t$, by means of time-slicing of the path-integral over particle 
configurations on $T^\ast T[1]M$, can be recovered by gauge fixing of a 
one-dimensional AKSZ sigma model 
with target
\begin{align}\label{eq:targetspace}
    T^*\big(T[1]M \times \mathbb{R}[1]\big)\,.
\end{align}

In \cite{Gozzi:2000sf}, the authors worked out all the different transformations 
of the fields appearing in the path integral formulation of classical mechanics, 
which we will briefly discuss in \cref{sec:CPI}, to show that they form the 
cotangent bundle $T^\ast T[1]M$ of the reversed-parity tangent bundle of the phase space $M$. 
In our case, we show that this identification is quite natural from the viewpoint 
of an one dimensional sigma model, opening up the possibility of generalizing the 
construction by replacing $\mathbb{R}$ in \cref{eq:targetspace} with a general group 
$G$. Most importantly, our observation would facilitate novel ways of looking at the 
symmetries and conservation laws of classical mechanics from the viewpoint of the rich 
geometric and topological structure of the AKSZ sigma models. The possibility of a 
connection between KvN formalism and geometric quantization as envisaged previously in 
\cite{Abrikosov:2003ce, Abrikosov:2004cf} can now be rigorously understood from the 
vantage point of AKSZ sigma models.

Our plan for this paper is the following.
In Section \ref{sec:CPI}, we begin with a brief review of the 
Koopman--von Neumann (KvN) formulation 
\cite{Koopman:1931, von_Neumann:1932i, von_Neumann:1932ii} of classical mechanics, as well its path integral 
reformulation due to Gozzi, Reuter and Thacker (GRT) in 
\cite{Gozzi:1989bf, Gozzi:1991di,Gozzi:2000sf}. 
Section \ref{sec:worldline} will then elaborate on the AKSZ action 
for the worldline of a particle. Finally in Section \ref{sec:toGRT} 
we show the exact connection between AKSZ 
sigma model and the GRT model and conclude with a summary of our results 
and future outlook in Section \ref{sec:discussion}.

\section{KvN and Classical Path Integral formulation}
\label{sec:CPI}
Classical mechanics and quantum mechanics are developed
on the basis of two completely different mathematical paradigms.
Contrary to the geometric approach of classical mechanics,
the description of quantum mechanics is more algebraic
in nature. A state in classical mechanics can be viewed 
as a point on a symplectic manifold, the phase space,
which is by definition endowed with a Lie bracket,
the Poisson bracket. Any individual observable
can then be described as some real-valued function
on this symplectic manifold, associated with a Hamiltonian vector field, 
generating individual flows on this manifold. For example, the flow corresponding 
to the Hamiltonian $H$ would describe the time evolution of the system. 
On the opposite side, the algebraic language of quantum mechanics revolves 
around the construction of a Hilbert space. 
Each physical 
state is described by a ray in the Hilbert space, and observables are the 
self-adjoint linear operators defined on the Hilbert space.
The Lie algebra structure appears by taking commutators
between different observables, i.e. it comes through
via the associative product defined by the composition
of operators acting on the Hilbert space.

During the era of 1930's several attempts were made to reconcile these two languages. 
Perhaps with this early motivation Koopman and von Neumann reformulated classical 
mechanics to associate it with a Hilbert space of complex and square-integrable 
functions similar to its quantum mechanical counter parts. Analogous to quantum 
mechanics one can also associate complex classical wave functions to a classical 
mechanical system of course with some caveats on which we will not focus in the current 
context and instead refer to \cite{Mauro:2003}.

Without loosing any generality, let us start with an one dimensional 
system with phase space density $\rho(q,p,t)$, which can be interpreted as 
the the probability of finding a particle at point $q$ with momentum $p$ 
exactly at time $t$ with the measure $\int dq\, dp$. Liouville theorem states 
that this density has the same property as an incompressible fluid, such that 
the phase space volume remains constant and the continuity equation becomes 
\begin{equation}
	\frac{d\rho}{dt} = \frac{\dow \rho}{\dow t}
	+ \dot{q}\frac{\dow \rho}{\dow q}
	+ \dot{p}\frac{\dow \rho}{\dow p} = 0\,.
\end{equation}
One can now use Hamilton's equations 
\begin{equation}
	\dot{q} = \frac{\dow H}{\dow p}\,,
	\quad \dot{p} = -\frac{\dow H}{\dow q}\,,
\end{equation}
to show that 
\begin{equation}\label{eq:conti}
	\frac{\dow\rho}{\dow t} = -\frac{\dow H}{\dow p}\,
	\frac{\dow \rho}{\dow q} + \frac{\dow H}{\dow q}\,
	\frac{\dow \rho}{\dow p}\,.
\end{equation}
One can easily generalise the discussion to a dynamical system with 
$n$ degrees of freedom in configuration space, corresponding to a phase space 
of dimension $2n$. 
Defining the Liouville operator as 
\begin{equation}
	\hat{\cL}=-i\,\frac{\partial H}{\partial p_i}\,\frac{\partial}{\partial q^i}
    +i\,\frac{\partial H}{\partial q^i}\,\frac{\partial }{\partial p_i}
\end{equation}
one can rewrite \eqref{eq:conti} as
\begin{equation}\label{eq:liouville}
	i\,\frac{\dow \rho}{\dow t} = \hat{\cL}\,\rho\;.
\end{equation}
In the following, we will use the notation 
$\boldsymbol{z}=(z^a)=(q^1, \ldots, q^n, p_1, \ldots, p_n)\,$ for 
the dynamical variables in phase space.
The basic postulates of Koopman and von Neumann formalism are
\begin{itemize}
	\item[(1)] the existence of a complex function $\psi(\boldsymbol{z},t)$ which obeys 
    the same dynamical equation as $\rho(\boldsymbol{z},t)$, i.e.,
	\begin{equation}\label{eq:post1}
		i\,\frac{\dow\psi(\boldsymbol{z},t)}{\dow t}=\hat{\cL}\,\psi(\boldsymbol{z},t)\;;
	\end{equation}
	\item[(2)] the definition of a scalar product
	\begin{equation}
		\cor{\psi}{\varphi}_t=\int d^{2n}z\,\,
        \psi(\boldsymbol{z},t)^*
        \varphi(\boldsymbol{z},t)\;.
	\end{equation}
\end{itemize}

Equation \eqref{eq:post1} can then be thought of
being the analogue of Schr\"odinger's equation
in quantum mechanics. The Hilbert space spanned by
the functions $\psi(\boldsymbol{z},t)$ can then be considered
as the Hilbert space for classical mechanics. 
The postulate of the scalar product ensures a proper definition
of the Hilbert space and imposes the norm squared of the states
to be
\begin{equation}
    \cor{\psi}{\psi}=\int d^{2n}z\,\,
    \psi(\boldsymbol{z},t)^*\psi(\boldsymbol{z},t)\;.
\end{equation}
With this definition of scalar product, one can further show
that $\hat{\cL}$ is an Hermitian operator%
\footnote{Assuming $\psi(\boldsymbol{z},t)$ to behave
in such a way that $\psi(\boldsymbol{z},t)|_{\boldsymbol{z}\rightarrow\infty}=0$.}
such that
\begin{equation}
\cor{\psi}{\hat{\cL}\,\varphi}=\cor{\hat{\cL}\,\psi}{\varphi}\,.
\end{equation}
%
The Hermitian character of $\hat{\cL}$ ensures
that $\cor{\psi}{\psi}$ remains conserved during the evolution.
Therefore one can now consistently interpret
\begin{equation}
    \psi^*(\boldsymbol{z},t)\psi(\boldsymbol{z},t)
    = \rho(\boldsymbol{z},t)
\end{equation}
as the density probability function,
and note that the Liouville theorem \eqref{eq:liouville} 
can be derived starting from the postulate \eqref{eq:post1}
of KvN mechanics itself. As evident from \eqref{eq:liouville}, 
although the classical wave function $\psi(\boldsymbol{z},t)$
is complex, the evolution of its phase is completely 
independent from its modulus, unlike the situation
in quantum mechanics. We will keep the implications
of this observation and further comparisons of KvN mechanics
and quantum mechanics for the excellent review
in \cite{Mauro:2003}, and move on to the path integral approach. 
More details of the following discussion can be found
in \cite{Gozzi:2000sf, Mauro:2003}.

In \cite{Gozzi:1986ge}, the author prescribes a simple way
to introduce a path integral formulation for classical mechanics. 
Unlike quantum mechanics, where each path is weighted
by a probability $\exp(\frac{i}{\hbar} S)$ with $S$ being the 
action of the path considered, in classical mechanics only the 
classical path between two fixed end points is allowed to have 
weight $1$, while all the others are weighted to zero. 
Nevertheless, Hamilton's variational principle 
considers all these virtual paths as well, only one being realised 
in the classical world as the one that extremises the action.
One can think of the classical analogue of the propagator, 
i.e., the probability of finding a classical particle at a point 
$\boldsymbol{z}$ in phase space at some time $t$, 
if it was initially at the point $\boldsymbol{z}_0$ at the time $t_0$:
\begin{equation}\label{classicalproba}
    K(\boldsymbol{z},t|\boldsymbol{z}_0^a,t_0)
    =\delta^{2n}\big(z^a-z_{cl}^a(t;\boldsymbol{z}_0,t_0)\big)\,,
\end{equation}
where by $z_{cl}^a(t;\boldsymbol{z}_0,t_0)$
we denote the classical solution of the Hamiltonian equations
of motion
\begin{equation}\label{eq:HamiltonEOM}
    \dot{z}^a=\pi^{ab}\dow_bH\;,
\end{equation}
given the initial condition $\boldsymbol{z}|_{t=t_0}=\boldsymbol{z}_0^a\,$, 
where $\pi=(\pi^{ab})$ is the inverse of the symplectic matrix. 

Slicing up the time interval $[t,t_0]$ into $N+1$ equal intervals $\delta t$,
and denoting the time in each interval as $t_i$
with $\boldsymbol{z}_i=\boldsymbol{z}(t_i)\,$ and $t_{N+1}=t\,$,
one can write the delta distribution in \eqref{classicalproba} as
\begin{equation}\label{eq:maindelta}
    \delta^{2n}\big(z^a-z_{cl}^a(t;\boldsymbol{z}_0,t_0)\big)
    = (\prod_{i=1}^N\int d\boldsymbol{z}_i)\, 
    \delta^{2n}\big(\boldsymbol{z}_{N+1}-\boldsymbol{z}_{cl}
    (t_{N+1};\boldsymbol{z}_N,t_N)\big)\ldots\,
    \delta^{2n}\big(\boldsymbol{z}_1-\boldsymbol{z}_{cl}
    (t_1;\boldsymbol{z}_0,t_0)\big)\,.
\end{equation}
Using \eqref{eq:HamiltonEOM} and having in mind the limit where $N\rightarrow \infty$, 
so that the interval $\delta t = t_{i+1}-t_i$ goes to zero, 
each of the delta distributions above 
($j\in\{0,1,\ldots,N\}$) can be re-written as
\begin{equation}
    \delta^{2n}\big(\boldsymbol{z}_{j+1}-\boldsymbol{z}_{cl}
    (t_{j+1};\boldsymbol{z}_j,t_j)\big)
    =\prod_{a=1}^{2n}\delta\big(\dot{z}^a
        -\pi^{ab}\dow_bH\big)|_{t=t_j}\;\, 
    \text{det}\left[\dow_t\delta^a_b
        -\pi^{ac}\dow_c\dow_bH\right]|_{t=t_j}\;,
\end{equation}
where we have made use of the standard formula 
$\delta^{2n}(z^a-z_{*}^a) = \delta^{2n}(g^a(\boldsymbol{z}))
\big|\det(\frac{\partial g^a}{\delta z^b})\big|$
where $g^a(\boldsymbol{z})
=\delta t\big(\frac{z^a_{i+1}-z^a_i}{\delta t}
-\pi^{ac}\partial_cH(\boldsymbol{z}_i)\big)\,$.
At this level of formality, 
the absolute value of the determinant is dropped.
Collecting all these definitions together
and taking the $N\rightarrow\infty$ limit%
\footnote{The normalization factor arising
due to this limit can be absorbed
into the path integral measure $\mathcal{D}{\cal Z}$.},
one can rewrite 
\begin{equation}\label{finalpath}
    \delta^{2n}\big(z^a-z_{cl}^a(t;\boldsymbol{z}_0,t_0)\big)
    = \int_{\boldsymbol{z}_0}^{\boldsymbol{z}}\mathcal{D}{\cal Z}\;
     \tilde{\delta}(\dot{\boldsymbol{z}}^a - \pi^{ab}\dow_bH)\,
    \text{det}(\dow_t\delta^a_b-\pi^{ac}\dow_c\dow_bH)\;,
\end{equation}
in a form of path integral in phase space,
where the symbol $\tilde{\delta}$ indicates a functional definition
for the product of the infinite number of delta function coming from  
\eqref{eq:maindelta} in the limit $N\rightarrow \infty\,$. 

One can then exponentiate both factors under the path integral 
in \eqref{finalpath} by introducing $2n$ variables $\lambda_a$ 
and a total of $4n$ anti-commuting variables $(\bar{c}_a,c^a)$ 
through the simple relations 
\begin{equation}
    \begin{split}
        \tilde{\delta}(\dot{z}^a-\pi^{ab}\dow_bH)
        &=\int \mathcal{D}\lambda\,\exp\left[i\int_{t_0}^t \dR t'\,
        \lambda_a(t')\big(\dot{z}^a
        -\pi^{ab}\dow_b H\big)\right]\;,\\
        \text{det}(\dow_t\delta^a_b
        -\pi^{ac}\dow_c\dow_bH)
        &=\int \mathcal{D}c\mathcal{D}\bar{c}
        \exp\left[\int_{t_0}^tdt'\bar{c}_a(t')
        (\dow_{t'}\delta^a_b-\pi^{ac}\dow_c\dow_bH)c^b(t')\right]\;.
    \end{split}
\end{equation}
The final result is that the propagator
in classical mechanics can be represented as the path-integral 
\begin{equation}\label{actionGRT}
    K(\boldsymbol{z},t|\boldsymbol{z}_0,t_0)
    =\int_{\boldsymbol{z}_0}^{\boldsymbol{z}} 
    \mathcal{D}{\cal Z}\mathcal{D}\lambda
    \mathcal{D}c\mathcal{D}\bar{c}\,
    \exp\left[i\int_{t_0}^tdt'\,{\cal L}\right]\,,
\end{equation}
with the Lagrangian ${\cal L}$ being
\begin{equation}\label{eq:LagGRT}
    {\cal L}=\lambda_a\dot{z}^a + i\bar{c}_a\dot{c}^a
    -\lambda_a\pi^{ab}\dow_bH
    -i\bar{c}_a\pi^{ad}(\dow_d\dow_bH)c^b\,.
\end{equation}
The first two terms provide one with a symplectic structure. 
The rest gives the extended Hamiltonian $\cal{H}$ as \cite{Gozzi:2000sf}, 
\begin{equation}
    {\cal{H}} = \lambda_a\pi^{ab}\dow_bH
    + i\bar{c}_a\pi^{ad}(\dow_d\dow_bH)c^b.
\end{equation}
Hence, starting from the original $2n$ dimensional phase space
with coordinates $z^a\,$, one arrives at an $8n$ dimensional 
extended phase space with coordinates $(z^a,\lambda_a,c^a,\bar{c}_a)$
where each of the paths in the path integral formulation
is weighted by a factor of 
$\exp[\,i\tilde{S}\,]=\exp[i\int dt\, \mathcal{L}]$, 
which by construction reproduces all the standard results
of classical mechanics. In the series of work pioneered
by E. Gozzi in \cite{Gozzi:1986ge}, the authors have explicitly 
searched for the geometric meaning of this $8n$ dimensional space 
which at this point seems like an abstraction
over the usual notions of the symplectic formulation
of classical mechanics. In the following sections
we will show how this apparent abstraction
of the extended $8n$ dimensional phase space can be understood
through a one-dimensional AKSZ sigma model.
Together with the equations of motion derived from $\mathcal{L}$
\begin{equation}
    \begin{split}
        \dot{z}^a&=\pi^{ab}\dow_bH,\\
        \dot{c}^a&=\pi^{ac}\dow_c\dow_bH\,c^b,\\
        \dot{\bar{c}}_b&=-\bar{c}_a\,\pi^{ac}\dow_c\dow_bH,\\
        \dot{\lambda}_b&=-\pi^{ac}\dow_c\dow_bH\,\lambda_a
        -i\,\bar{c}_a\,\pi^{ac}\dow_c\dow_d\dow_bH\,c^d,
    \end{split}
\end{equation}
and the transformations of each of these new fields
under symplectic diffeomorphisms of $z^a$,
the authors in \cite{Gozzi:2000sf, Mauro:2003}
correctly concluded that the phase space spanned by the $8n$ variables 
$(z^a,\lambda_a,c^a,\bar{c}_a)$ is $T^\ast T[1]M$,
where $M$ is the symplectic manifold coordinatised by the original
$2n$ variables $z^a$. We remark that here and in the rest of the paper, 
we work in Darboux coordinates, only allowing for canonical transformations
instead of all the possible diffeomorphisms of $M$. 

\section{Worldline model: AKSZ approach}
\label{sec:worldline}

In this section, we show that the identification of 
a phase space $T^\ast T[1]M$ automatically 
follows from the AKSZ treatment of a classical particle.

\paragraph{Constraints from Lie algebra actions.}
Instead of considering an unconstrained particle evolving though 
an Hamilton flow, as we did in the previous section, 
here we consider a constrained particle whose phase space corresponds
to a symplectic manifold $(\M,\omega)$, subject to a system
of first class constraints $T_I$ which define a representation
of a Lie algebra $\g$ on the algebra of functions $\Functions(\M)$.
In other words, we assume that the constraints $T_I$
assemble into an equivariant moment map
$T:\M \longrightarrow \g^*$,
meaning that the functions $T_I$ are obtained as
$T_I(x) := \langle T(x), \mathsf{t}_I \rangle_\g$
for $x \in \M$ and $\{\mathsf{t}_I\}$ a basis of $\g$, 
and where $\langle-,-\rangle_\g$ denotes the canonical
pairing between $\g$ and its linear dual $\g^*$).
Such a system can be encoded in the degree $0$
symplectic $\cQ$-manifold $\M \times T^*\g[1]$, 
with cohomological vector field
\begin{equation}\label{eq:BFV}
	\cQ := \pmb\{\Theta,-\pmb\}
	\qquad \text{where} \qquad
	\Theta = c^I\,(T_I
	- \tfrac12\,f_{IJ}{}^K\,c^J\, \mathcal{P}_K)\,,
\end{equation}
and where $\{c^I\}$ are degree $+1$ coordinates
on $\g[1]$ and $\{\mathcal{P}_I\}$ are their momenta,
i.e. degree $-1$ coordinates in the fiber directions
of $T^*\g[1]$, and $\pmb\{-,-\pmb\}$ is the sum of
the Poisson brackets on $\M$ and the canonical Poisson
bracket on $T^*\g[1]$. The algebra of functions on
this graded manifold is isomorphic to the complex
\begin{equation}
	\wedge(\g \oplus \g^*) \otimes \Functions(\M)\,,
\end{equation}
equipped with the differential
\begin{equation}
	\cQ = \delta_\g + \delta\,,
\end{equation}
where 
\begin{equation}
	\delta_\g = c^I\,\Big(\{T_I,-\}
	+ f_{IJ}{}^K\,\mathcal{P}_K\,
	\tfrac{\partial}{\partial \mathcal{P}_J}
	-\tfrac12\,f_{IJ}{}^K\,c^J\,
	\tfrac{\partial}{\partial c^K}\Big)\,
\end{equation}
is the Chevalley--Eilenberg differential on the module
$\wedge \g \otimes \Functions(\M)$, and
\begin{equation}
	\delta = T_I\,
	\tfrac{\partial}{\partial \mathcal{P}_I}\,
\end{equation}
is the Koszul differential. In other words, it is nothing but
the Batalin--Fradkin--Vilkovisky (BFV) 
\cite{Fradkin:1977hw,Batalin:1977pb,Fradkin:1977xi} and 
Batalin--Vilkovisky (BV) \cite{Batalin:1981jr,Batalin:1983ggl} 
extensions of the Becchi--Rouet--Stora--Tyutin (BRST) 
\cite{Becchi:1974md,Becchi:1975nq,Tyutin:1975qk}
complex associated with the constrained system
described by $(\M, \{T_I\})$. Note that AKSZ sigma-model
having the BFV--BRST phase space of a constrained system
were first introduced and discussed in \cite{Grigoriev:1999qz}.
For reviews on the BV, BFV, and BRST approaches
to gauge theories, see e.g. \cite{Henneaux:1992ig, Gomis:1994he, Fuster:2005eg, Jurco:2018sby, Barnich:2018gdh, Cattaneo:2019jpn, Cattaneo:2023hxv}.

\paragraph{AKSZ action.}
Let us assume that the symplectic form on $\M$ is exact,
i.e., that we can write
\begin{equation}
	\omega = \dR \vartheta\,,
\end{equation}
for some one-form $\vartheta\in\Omega^1(\M)$. 
The source manifold $\Sigma$ of the sigma-model
we shall consider is the world-line of the classical particle. 
One can write the AKSZ action associated with
$\M \times T^*\g[1]$ in terms of ``super-maps''
\begin{equation}
	T[1]\Sigma \longrightarrow \M \times T^*\g[1]\,,
\end{equation}
whose components give rise to the ``super-fields''
\begin{equation}
	\boldsymbol{x}^\mu := x^\mu + \theta\,\pi^{\mu\nu}(x)\,x_\nu^+\,,
	\qquad
	\boldsymbol{e}^I := c^I + \theta\,e^I\,,
	\qquad
	\boldsymbol{c}_I := e^+_I + \theta\,c^+_I\,,
\end{equation}
where $\theta$ is the odd coordinate on $T[1]\Sigma$
and the bivector $\pi^{\mu\nu}(x)$ is the inverse
of the symplectic form $\omega_{\mu\nu}$ of $\M$.
The classical fields, of ghost number $0$, consists of
a map $x: \Sigma \to \M$ from the worldline $\Sigma$
to the target space $\M$  and a Lagrange multiplier
$e = e^I\,\mathsf{t}_I$ which takes values in $\g$.
It can be thought of as an einbein, or a gauge field
on the worldline $\Sigma$, whose gauge parameters
give rise to the ghosts $c=c^I\,\mathsf{t}_I$,
of ghost number $1$ and $\g$-valued.
The corresponding antifields $x^+$, $e^+$ and $c^+$
are of ghost number $-1$, $-1$ and $-2$ respectively.

The AKSZ action is then given by
\begin{equation}
	S_{\sf\sst AKSZ}[\boldsymbol{x}, \boldsymbol{e}, \boldsymbol{c}]
	= \int_{T[1]\Sigma} (\vartheta_\mu(\boldsymbol{x})\,
	\boldsymbol{d}_\Sigma \boldsymbol{x}^\mu
    + \boldsymbol{c}_I\,\pmb{d}_\Sigma \boldsymbol{e}^I
	- T_I(\boldsymbol{x})\, \boldsymbol{e}^I
	+ \tfrac12\, f_{IJ}{}^K\, \boldsymbol{e}^I\,\boldsymbol{e}^J\,
    \boldsymbol{c}_K)\,,
\end{equation}
where 
\begin{equation}
    \boldsymbol{d}_\Sigma = \theta\,\tfrac{\partial}{\partial\tau}\,,
\end{equation}
is the homological vector field corresponding
to the de Rham differential on $\Sigma$, and the integration
over $T[1]\Sigma$ is defined as
\begin{equation}
    \int_{T[1]\Sigma}(-) = \int_\Sigma \dR\tau \int \dR\theta(-)\,,
\end{equation}
with the piece $\int\dR\theta(-)$ being the Berezin integral.
In the AKSZ scheme, selecting the top-form component amounts to 
restricting to the ghost number zero piece of the integrand.
Performing the Berezin integral yields
\begin{equation}
    \begin{aligned}
    	S_{\sf\sst AKSZ}[x,e,c] & = \int_\Sigma \dR\tau\,
    	\Big(\vartheta_\mu(x)\,\dot x^\mu
    	- \langle T(x), e \rangle_\g \\
        & \hspace{80pt} -x_\mu^+\,c^I\,\{T_I,x^\mu\}
    	+ \langle e^+, \dot c + [e,c] \rangle_\g 
    	+ \tfrac12\,\langle c^+, [c,c] \rangle_\g\Big)\,,
    \end{aligned}
\end{equation}
where we write $\tau$ for the worldline coordinate,
the dot over a field denotes its derivative with respect
to $\tau$, e.g. $\dot x^\mu = \tfrac{\dR x^\mu}{\dR \tau}$,
and $\langle-,-\rangle_\g$ denotes the pairing between
$\g$ and $\g^*$. The classical piece of this action,
obtained by setting all fields of ghost number different
than zero, is simply given by
\begin{equation}
	S_{\sf\sst cl.}[x,e]
	= \int_\Sigma \vartheta_\mu(x)\,\dot x^\mu
	- e^I\,T_I(x)\,,
\end{equation}
and is invariant (modulo a boundary term) under
the gauge symmetry 
\begin{equation}
	\delta_\epsilon x^\mu = \epsilon^I\, \{T_I,x^\mu\}\,,
	\qquad
	\delta_\epsilon e^I = \dot\epsilon^I
	+ f_{JK}{}^I\, e^J\,\epsilon^K\,,
\end{equation}
with the gauge parameter $\epsilon \in \Functions(\Sigma,\g)$.
The corresponding equations of motion read
\begin{equation}
	\dot x^\mu = e^I\,\{T_I,x^\mu\}\,,
	\qquad\qquad
	T_I = 0\,,
\end{equation}
In plain words, the trajectories are constrained
on the surface defined by the zeroes of the moment map,
and evolve along the fundamental vector fields
of the $\g$-action.

\paragraph{Gauge fixing.}
Let us single out a direction in $\g$,
denoted by the value $\bullet$ and choose the gauge
wherein only this component of the einbein is fixed to $1$
and all others to zero. To do so, we should add
the non-minimal sector to the AKSZ action,
\begin{equation}
	S_{\smf non-min.}[b,\bar c^{\,+}]
	= \int_\Sigma \dR\tau\,b_I\,\bar{c}^{\,+I}\,,
\end{equation}
where both $b_I$ and $\bar c^{\,+I}$ have ghost number $0$,
and encode this choice of gauge through the gauge fixing fermion
\begin{equation}
	\Psi[\bar c, e]
	= \int_\Sigma \dR\tau\,\bar c_I\,(e^I - \delta^I_\bullet)\,.
\end{equation}
Its variation with respect to each field fixes
the value of the corresponding antifield, 
which in our case yields
\begin{equation}
	\bar c^{\,+I} = e^I - \delta^I_\bullet\,,
	\qquad 
	e^+_I = \bar c_I\,,
	\qquad 
	x^+_\mu = 0 = c^{+I}
\end{equation}
so that the gauge fixed action becomes
\begin{equation}
	S_{\smf g.f.}[x,e,c,\bar c,b]
	= \int_\Sigma \dR\tau\Big[\vartheta_\mu(x)\,\dot x^\mu
	- e^I\,T_I(x) + \bar c_I\,(\dot c^I + f_{JK}{}^I\,e^J\,c^K)
	+ b_I\,(e^I-\delta^I_\bullet)\Big]\,.
\end{equation}
Upon integrating out the Lagrange multiplier $b_I$, 
one finds
\begin{equation}
	S_{\smf g.f.}[x,c,\bar c] = \int_\Sigma \dR\tau\,
	\Big[\vartheta_\mu(x)\,\dot x^\mu - \cH(x)
	+ \bar c_I\,(\dot c^I + \rho_J{}^I\,c^J)\Big]\,,
\end{equation}
where we re-named $T_\bullet(x) \equiv \cH(x)$
and $f_{\bullet\,J}{}^I \equiv \rho_J{}^I$.

\paragraph{Generic first class constraints.}
Suppose that the constraints $T_I$ do not
come from a moment map for some Lie algebra,
but are instead generic first class constraints,
\begin{equation}
	\{T_I, T_J\} = C_{IJ}{}^K\,T_K\,,
\end{equation}
where $C_{IJ}{}^K \equiv C_{IJ}{}^K(x)$
are structure \emph{functions}, i.e. they may depend
non-trivially on the phase space coordinates $x^\mu$.
In this more general case, the target space BRST charge
$\Theta$ receives corrections,
\begin{equation}
	\Theta(x,c,\mathcal{P}) = c^I\,T_I
	- \tfrac12\,C_{JK}{}^I\,c^I c^J\mathcal{P}_K + \dots\,,
\end{equation}
where the dots denote term of higher order
in ghost momenta $\mathcal{P}$.
We can nevertheless write the corresponding AKSZ action
\begin{align}
	S_{\sf\sst AKSZ}[x,e,c] & = \int_\Sigma \dR\tau\,
	\Big[\vartheta_\mu(x)\,\dot x^\mu + e^+_I\,\dot c^I
	- x_\mu^+\pi^{\mu\nu}\,
	\tfrac{\delta}{\delta x^\nu}\Theta(x,c,e^+) \\
	& \hspace{100pt}
	- e^I\,\tfrac{\delta}{\delta c^I}\Theta(x,c,e^+)
	- c^+_I\,\tfrac{\delta}{\delta e^+_I}\Theta(x,c,e^+)\Big]\,.
\end{align}
The same gauge fixing as in the previous paragraph
can be implemented, to give
\begin{equation}
	S_{\smf g.f.}[x,e,c,\bar c,b]
	= \int_\Sigma \dR\tau\Big[\vartheta_\mu(x)\,\dot x^\mu
	- e^I\,\tfrac{\partial}{\partial c^I}\Theta(x,c,\bar c)
	+ \bar c_I\,\dot c^I + b_I\,(e^I-\delta^I_\bullet)\Big]\,,
\end{equation}
which can be further simplified by integrating out $b_I$,
\begin{equation}
	S_{\smf g.f.}[x,e,c,\bar c,b]
	= \int_\Sigma \dR\tau\Big[\vartheta_\mu(x)\,\dot x^\mu
	- \tfrac{\delta}{\delta c^\bullet}\Theta(x,c,\bar c)
	+ \bar c_I\,\dot c^I\Big]\,.
\end{equation}

\section{Recovering the GRT formulation}
\label{sec:toGRT}

In order to make contact with the Lagrangian \eqref{eq:LagGRT}
and the other results of \cite[Sec. 3]{Gozzi:1989bf} reviewed in 
Section \ref{sec:CPI}, 
it appears that one must consider the cotangent bundle of the phase space
of our original system, i.e. $\M=T^*M$
where $M$ is the original symplectic manifold with local coordinates 
$(z^a)_{a=1, \ldots, 2n}\,$, as this would account
for the classical fields $(z^a, \lambda_a)$
where $\lambda_a$ are conjugated to $z^a$ in $T^*M$.
In other words, we have the decomposition 
\begin{equation}
(x^\mu)_{\mu=1,\ldots,4n } = (z^a, \lambda_a)_{a=1,\ldots,2n}\,.    
\end{equation}
On top of that, the symplectic potential on $\M$ is taken 
to be canonical,
$\vartheta = \vartheta_{\mu}(x)\dR x^\mu = \lambda_a\,\dR z^a$, 
which does lead to the kinetic term $\int_\Sigma \dR\tau\,\lambda_a \,\dot z^a\,$.
To account for the interactions in the action,
as a result of a gauged fixed AKSZ action as described above,
we find that one should use a BFV--BRST charge of the form
\begin{equation}
    \Theta(z,\lambda,c,\mathcal{P})
    = c^\bullet\Big(\lambda_a\,\pi^{ab}\,\partial_b H(z)
    + c^b\, \mathcal{P}_a\,\pi^{ac}\,
    \partial_c \partial_b H(z)\Big) + \dots\,,
\end{equation}
where we recall the notation
$\partial_a \equiv \tfrac{\partial}{\partial z^a}$,
and where the dots denote terms which are independent
of $c^\bullet$ and which ensure that
$\pmb\{\Theta,\Theta\pmb\} = 0$.

This suggests that the constrained system 
described by the sought-for BRST charge $\Theta$ 
is determined by chosing the constraints 
\begin{equation}
 T_I = (T_\bullet\,, \,T_a)\;,\quad i.e.\quad I=(\bullet\,,a)\;,
\end{equation}
where
\begin{equation}
	T_\bullet = \lambda_a X^a_H(z)\,,
	\qquad 
	T_a = \lambda_a\,,
	\qquad\text{with}\qquad
	X^a_H(z) = \{z^a,H\}_M
	= \pi^{ab}\,\partial_b H(z)\,,
\end{equation}
which verify
\begin{equation}
	\{T_\bullet, T_a\}_{T^*M} = C_{\bullet a}{}^b\,T_b\,,
	\qquad\text{with}\qquad
	C_{\bullet a}{}^b = \partial_a X^b_H\,.
\end{equation}
These constraints are all first class,
in accordance with our working assumption,
and the structure functions are nothing but
the first derivatives of the components
of the Hamiltonian vector field of the Hamiltonian $H$.
A direct computation shows that the BRST charge
\begin{align}
	\Theta(z,\lambda,c,\mathcal{P})
	& = c^\bullet\,T_\bullet + c^a\,T_a
	- C_{\bullet a}{}^b\,c^\bullet c^a\,\mathcal{P}_b
    \nonumber \\
    & = c^a\,\lambda_a
    + c^\bullet\,\pi^{ab}\,\lambda_a\,\partial_b H(z)
	- c^\bullet c^b\,\pi^{ac}\,
    \partial_b\partial_c H\,\mathcal{P}_a\,,
    \label{eq:finalBFVforGRT}
\end{align}
with the previously defined constraints
and structure functions, does indeed satisfy
$\pmb\{\Theta,\Theta\pmb\}=0$.

Therefore, we have shown that the action of \cite[Sec. 3]{Gozzi:1989bf}
reproduced in the exponential in Eq. \eqref{actionGRT}
is recovered from a gauge fixed AKSZ model
in one dimension, whose target space is associated
with the system of first class constraints
$\{T_\bullet\,,\,T_a\}$ on $T^*M$. 
This is the main result of the present paper.

Let us discuss these constraints. The easiest ones are $T_a=\lambda_a$
which identify the constraint surface as a submanifold
of the phase space $T^*M$. In fact, $T_\bullet$ does not
specify further the constraint surface, as it vanishes 
already on $M \equiv \{(z^a,\lambda_a=0)\} \subset T^*M$.
Recall however that in the presence of first class
constraints, one is interested in
the reduced phase space, that is, the quotient
of the constraint surface by the action of the distribution
generated by the first class constraints.
This is where $T_\bullet$ becomes relevant for us,
as quotienting $M$ by its action yields
the set of classical trajectories (the flows generated by
the Hamiltonian $H$) as the reduced phase space of our model.

\paragraph{Constraints from the shifted tangent bundle.}
Let us re-derive this constrained system
from a different perspective. Suppose we are given
a symplectic manifold $M$ and an Hamiltonian
$H \in \Functions(M)$.
The latter defines an action on $M$ of $\R$ viewed a Lie group with 
addition as its multiplication rule,
whose fundamental vector field is thus the associated
Hamiltonian vector field $X_H = \{H,-\}$. 
The integral curves
of this vector field, which are nothing but
the classical trajectories of this mechanical system,
correspond to the orbits of $\R$ on $M$.
Therefore the set of classical solutions 
can be identified with the quotient $M/\R$, 
the set of the aforementioned orbits.

In order to recover the space of classical solutions
from a one-dimensional AKSZ sigma model, to be identified 
with the previous model, we should find
a BFV description of this space, i.e., identify the symplectic 
$\cQ$-manifold of degree $0$ encoding the space
of classical trajectories as the result
of a coisotropic Weinstein reduction. 
In other words, let us look for a constrained system, 
with only first class constraints,
such that the orbits of the gauge symmetry generated
by the latter on the constraint surface, is isomorphic
to the set of classical trajectories.

We have recalled that the space of classical solutions
can be thought of as the set of orbits of the $\R$-action
generated by the Hamiltonian $H$, on the phase space $M$.
So we should find a way to recover the latter as a constraint
surface, in another symplectic manifold. One simple manner
to do so is to consider the cotangent bundle $T^*M$,
with coordinates $(z^a,\lambda_b)$ where $z^a$
are coordinates on $M$ and $\lambda_a$ the associated momenta,
i.e., the coordinates along the fibres of $T^*M$.
The original manifold $M$ can then be recovered as
a constraint surface defined by 
\begin{equation}
	M = \big\{(z^a,\lambda_a) \in T^*M \mid \lambda_a=0\big\}\,,
\end{equation}
or more geometrically, by identifying $M$ as the zero section
\begin{equation}
	\zeta:\ M \longhookrightarrow T^*M\,, 
\end{equation}
of its cotangent bundle $T^*M \twoheadrightarrow M$.
We can lift the $\R$ action to $T^*M$, where it becomes
Hamiltonian, generated by the cotangent lift of $X_H$.
To summarise, this reasoning leads us to considering
the same system of first class constraints
as we proposed before, that is
\begin{equation}
	T_a := \lambda_a\,,
	\qquad 
	T_\bullet = X_H^a\lambda_a
	= \pi^{ab}\,\lambda_a\,\partial_b H\,.
\end{equation}
The first ones, $T_a$, identify $M$ as the constraint surface
in $T^*M$, while the last one, $T_\bullet\,$, corresponds
to the $\R$-action lifted to $T^*M$.

At this point, we can make two observations. First,
the Hamiltonian vector fields associated with $T_a$
obviously form an integrable distribution, as they span
the tangent bundle of $M$ at any point, and hence
the Lie algebroid associated with it is simply $TM$.
Second, the $\R$-action also defines a Lie algebroid,
as any action of a Lie algebra on a manifold does,%
\footnote{Indeed, given a Lie algebra $\g$ which acts
	on a smooth manifold $M$ via
	$\rho: \g \longrightarrow \Gamma(TM)$, one can endow
	the trivial bundle $M \times \g \twoheadrightarrow M$
	with a structure of Lie algebroid as follow.
	First, notice that sections of this trivial bundle
	are nothing but $\g$-valued functions on $M$, i.e.
	$\Gamma(M \times \g) \cong \Functions(M) \otimes \g$.
	The anchor of the Lie algebroid is therefore
	simply given by the $\Functions(M)$-linear extension
	of the $\g$-action $\rho$. Explicitly, for a section
	$\psi \in \Gamma(M \times \g)$ written as
	$\psi = \psi^a(x)\,\mathsf{t}_a$ with $\{\mathsf{t}_a\}$
	a basis of $\g$, one defines the action of the anchor
	on it via
	\[
    	\rho_\psi := \psi^a(x)\,\rho_{\mathsf{t}_a}
    	= \psi^a(x)\,\rho_a{}^\mu(x)\partial_\mu\,.
	\]
	The Lie bracket is defined as 
	\[
    	[\psi_1,\psi_2]_{M\times\g}
    	= \big(\psi_1^a\rho_a{}^\mu\partial_\mu \psi_2^c
    	- \psi_2^a\rho_a{}^\mu\partial_\mu \psi_1^c
    	+ \psi_1^a \psi_2^b\,f_{ab}{}^c\big)\mathsf{t}_c\,,
	\]
	which can be thought of as a `twist' of the Lie bracket
	of $\g$ by the action of the latter on $\Functions(M)$.}
denoted $M \rtimes \R$, whose underlying vector bundle
is the trivial one, $M \times \R$, and which has only 
a non-trivial anchor in the guise of the fundamental
vector $X_H$ generating the action of $\R$.
Both $TM$ and $M \rtimes \R$ are Lie algebroids over $M$,
and hence so is their direct (or Whitney) sum,
that we shall denote by 
\begin{equation}
	E := TM \rtimes \R\,.
\end{equation}
Any Lie algebroid famously gives rise to a $\cQ$-manifold
\cite{Vaintrob:1997}, so in our case $E[1]$ is a graded manifold 
with coordinates $z^a$ of degree $0$, corresponding
to coordinates on $M$, and degree $1$ coordinates $c^a$
and $c^\bullet$ corresponding to coordinates along the fibres
of $TM$ and $M \rtimes \R$ respectively.
The cohomological vector field making $E[1]$
into a $\cQ$-manifold reads
\begin{equation}
	\cQ_E = c^a\,\tfrac{\partial}{\partial z^a}
	+ c^\bullet\,\pi^{ab}\,\partial_b H\,
	\tfrac{\partial}{\partial z^a}
	-c^\bullet\,c^b\,\pi^{ac}\,\partial_b\partial_c H\,
	\tfrac{\partial}{\partial c^a}\,,
\end{equation}
in this coordinate system. The cotangent bundle
of this $\cQ$-manifold
\begin{equation}
	T^*E[1] = T^*\big(T[1]M \rtimes \R[1]\big)\,,
\end{equation}
defines the symplectic $\cQ$-manifold
encoding the BFV description of the classical trajectories
we discussed, in accordance with the results and observations
of \cite{Ikeda:2018rwe}. Recalling from \eqref{eq:BFV}
that the momentum of $c^a$ is $\mathcal{P}_a$,
the cotangent lift of $\cQ_E$ is given by
\begin{equation}
	\Theta = c^a\,\lambda_a 
	+ c^\bullet\pi^{ab}\,\lambda_a\,\partial_b H
	- c^\bullet c^b \pi^{ac}\,
	\partial_b\partial_c H\,\mathcal{P}_a\,,
\end{equation}
that exactly reproduces the BFV--BRST charge
\eqref{eq:finalBFVforGRT} defining the AKSZ sigma model,
with target space $T^*E[1]$, and whose gauge fixing
reproduces the GRT one \cite{Gozzi:1989bf}, as we have shown.

\section{Conclusion}
\label{sec:discussion}
This paper offers a way to rethink classical mechanics
as a gauge fixed AKSZ sigma model. The way we showed this
is to start from the GRT reformulation of KvN classical mechanics.

In the GRT reformulation of classical mechanics, one considers
a simple classical system evolving in phase space
along a Hamiltonian flow. The authors of 
\cite{Gozzi:1989bf, Gozzi:1991di, Gozzi:2000sf}
proposed a path integral prescription for the classical system, 
for which the price to pay is the introduction
of additional fields. They show that, for the classical motion
of system with $n$ degrees of freedom in configuration space,
one has to introduce a total of $8n$ fields to ensure 
the consistency of the path integral and to reproduce
the expected classical trajectories in the phase space
of dimension $2n$. As observed in \cite{Gozzi:2000sf},
these $8n$ variables span $T^\ast T[1]M$ where $M$
is the $2n$ dimensional phase space for the system
under consideration. 

In this paper, we considered the worldline of a particle 
constrained by a set of first class constraints, and wrote
the AKSZ action corresponding to that constrained particle.
We showed that the gauge-fixed version of the AKSZ action,
for a suitable choice of target space and constraints
spelled out in Section \ref{sec:toGRT}, reproduces the action 
that dictates the GRT path integral formulation
of classical mechanics. 

We then reinterpreted our AKSZ sigma model as the  
BFV description of the constrained system that was designed  
to reproduce the GRT formulation of a classical, unconstrained, 
dynamical system in a phase space $M$. The reduced phase space
of this constraint system on $T^\ast M$, which consists of
the set of classical trajectories of the original
mechanical system encoded by $M$ and the Hamiltonian $H$,
is recovered by taking the quotient of $M$ by the distribution
associated with the Lie algebroid $TM \rtimes \R$.
In particular, the last factor $\R$ accounts for the flow
generated by the Hamiltonian $H$ of the original system.
These observations confirm our claim that a classical system, 
whose phase space corresponds to the symplectic manifold $M$,
is equivalent to a gauge fixed one-dimensional AKSZ sigma model 
with target space $T^\ast(T[1]M \times \R[1])$.

This simple yet intriguing mapping between a classical system
and a gauge-fixed  AKSZ sigma model opens up interesting avenues 
of research. One direct application of this mapping would be
to start with a constrained classical system, and look
for its AKSZ counter part. In particular, one could consider
the case of first class constraints generated by the action
of a Lie group $G$ on $M$, whose BFV--BRST description
leads to an AKSZ model with target space $T^\ast(M \rtimes \g[1])$,
where $\g$ is the Lie algebra of $G$. In light of the previous
treatment of an unconstrained classical system,
one could expect that the relevant target space be of the form
$T^\ast\big(T[1]C \rtimes (\R[1] \oplus \g[1])\big)$, 
where $C$ is the constraint surface defined by the first class constraints. 

As another direction of research, one can study higher dimensional 
sigma models and look for an effective classical mechanical system 
equivalent to it. Other interesting avenue would be to understand 
the connection between geometric quantization
(see \cite{Kostant:1974, Dudley:1970, Kirillov:2001}
for original references, and e.g., \cite{Woodhouse:1992, Wernli:2023pib})
and KvN mechanics
as shown by \cite{Abrikosov:2003ce,Abrikosov:2004cf}
in more details now in the light of AKSZ sigma models.

\section*{Acknowledgements} 
We acknowledge stimulating discussions with Per Sundell. 
One of the authors (A.C.) is grateful to Per Sundell for early collaboration.  
The work of T.B. was supported by the European Research Council 
(ERC) under the European Union’s Horizon 2020 research
and innovation programme (grant agreement No 101002551). 
The work of A.C. was partially supported by the European Union's Horizon 2020 research 
and innovation programme under the Marie Sk\l{}odowska Curie grant agreement number 
101034383.


\providecommand{\href}[2]{#2}\begingroup\raggedright\endgroup

\end{document}